# Spin-Polarisation measurement using NbN-Insulator-Ferromagnet Tunnel Junction with oxidized barrier

Pritam Das, John Jesudasan, Rudheer Bapat and Pratap Raychaudhuri[*]

*Tata Institute of Fundamental Research, Homi Bhabha Road, Mumbai 400005.*

We report a two-step process for the fabrication of superconductor-insulator-normal metal tunnel junctions using NbN as the superconducting electrode and its surface oxide as the insulating tunnel barrier, and investigate its efficacy in measuring spin-polarisation of ferromagnets using the Meservey-Tedrow technique. We observe that for NbN film thickness below 10 nm, under the application of parallel magnetic field, the superconducting density of states show clear "Zeeman" splitting into spin-up and spin-down sub-bands. Tunnelling measurements on devices in which ferromagnetic Co is used as the normal electrode show that these devices can be used to reliably measure spin polarisation of a ferromagnet at temperatures up to 1.6 K. The simplicity of our fabrication process, and the ability to perform spin-polarisation measurements at $^4$He temperatures make NbN a very attractive candidate for spin polarisation measurements.

---

[*]E-mail: pratap@tifr.res.in

**I. Introduction**

Determining the spin-polarization of ferromagnets is of central importance for their application in spintronic devices. Of the various methods used for measuring polarization[1,2,3,4,5,6], transport-based methods such as, point contact Andreev reflection[2,7] and superconducting tunneling spectroscopy (STS), are particularly popular due to their relative simplicity. Of these the STS technique pioneered in the 1970's by Meservey and Tedrow[4] can measure both the magnitude and sign of the spin-polarization in the tunneling current. This technique relies on splitting the superconducting density of states into spin-up and spin-down sub-bands by applying a large magnetic field parallel to the plane of a superconducting film[8,9], such that when the other tunneling electrode has a spin polarized conduction band, the spin-current preferentially flows from one spin band. Consequently, this creates an asymmetry between the spectral features associated with the two spin bands which can be analyzed to obtain the spin-polarization ($P$) of a ferromagnet.

In Meservey-Tedrow measurements, there are two important factors that can smear the spin-split density of states in the superconductor and limit the sensitivity of this technique: diamagnetic orbital current and spin-orbit coupling. To minimize orbital currents, the superconducting film thickness has to be much smaller than the London penetration depth ($\lambda$) and the coherence length ($\xi$), and the magnetic field needs to be precisely aligned along the film plane. Spin-orbit coupling, on the other hand, is a material property. The presence large spin-orbit coupling rule out the use of most of the elemental superconductors from MT measurements, with a handful of exceptions like Al and Ta. Al has been most widely used[10,11,12,13], due to high reproducibility of Al based tunnel junction. However the $T_c$ of Al films are low ( ~ 1.5 - 2.5 K ), which makes it necessary to perform measurements below 1 K using ³He cryostats. Therefore, a recent report of spin polarization measurement using NbN, a superconductor with much higher $T_c$ ( ~ 16 K ) is particularly interesting. Even though in very thin films the $T_c$ reduces from its bulk value, it still remains >11 K down to about 5 nm. However, on the downside the device fabrication was relatively complex, requiring four separate shadow masks to form sequentially the NbN bottom electrode, the MgO tunnel barrier, isolation pads, and finally the top electrode.[14]

In this paper, we present a simple two-step process using two shadow masks to fabricate NbN based tunnel devices. The device fabrication gets vastly simplified by using an oxidised surface layer of NbN as the tunnel barrier. First, we investigate NbN/oxide/Ag tunnel junctions with various thickness (*t*) of NbN. We show that signatures of spin splitting in the tunnelling spectra appears for $t \leq 10\ nm$ and becomes pronounced for $t \sim 5\ nm$. We also show that various materials parameters extracted from these spectra follow their expected variation from theory. Subsequently, fabricating NbN/oxide/Co tunnel junction we show that consistent and reproducible values of spin-polarisation of the ferromagnet can be obtained at temperatures up to 1.6 K.

## II. Experimental methods

Thin films of NbN were synthesized on (100) oriented single crystalline MgO substrate through reactive DC magnetron sputtering by sputtering a Nb target in an Ar-$N_2$ gas mixture keeping the substrate at $600^0$C. Details of NbN thin films growth have been reported elsewhere.[15] When depositing the NbN films we used the optimum conditions that gave us $T_c$ ~ 16 K for films with $t > 10\ nm$ and only varied the deposition time when synthesizing thinner films. We used a shadow mask to deposit the NbN film in the form of a 300 μm wide strip as shown in Fig. 1(a). Planar Superconductor-insulator-normal metal tunnel junction (SIN) was fabricated by oxidizing the NbN surface at 250 °C for 1 hr 30 min to 2hrs, either in atmosphere or in high purity oxygen for making the insulating tunnel barrier, and then depositing cross strip of Ag or Au by at room temperature (Fig. 1(a)). Cross sectional Transmission Electron microscope images showed that this oxidation protocol resulted in a uniform amorphous oxide layer with thickness ~ 2 nm (Fig. 1(b)). While the oxidised barriers formed in air or in pure oxygen showed no significant morphological difference, the tunnelling characteristics of the junction showed some differences. This will be discussed later. Our 6-terminal device configuration allowed us to measure the tunnel junction resistance and the resistance of the NbN film in the same run, by using two difference configuration for passing the current and measuring the voltage ( Fig. 1(c) ). Magneto-transport measurements are performed in a conventional $^3$He cryostat fitted with a 110 kOe superconducting solenoid. The sample is mounted on a sample rotator which is adjusted to align the magnetic field parallel to the film plane. The tunnelling conductance $\left(G(V) = \frac{dI}{dV}\Big|_V\right)$ is obtained by measuring the current-voltage (*I-V*) characteristics of the tunnel junction using a precision source and a nanovoltmeter and numerically differentiating the data.

The Superconductor-insulator-Ferromagnetic (SIF) tunnel junctions were made in the same way, but replacing the normal metal electrode with 10-20 nm thick cross strip of Co. One problem that we faced is that the resistance of the Cobalt strips is comparable to the resistance of the tunnel junction, and thus generates artefacts originating from series resistance of the Co arms leading to the contact pads. To eliminate this problem, we deposited an over-layer of Au on the Co strips, which provided a parallel low resistance path for the current to flow outside the tunnel junction.

**III. Results and Discussion**

III.A. *Spin splitting of the tunnelling density of states in SIN junctions*

In Fig. 2 (a)-(d) we show the $G(V)$ $vs. V$ for 4 NbN/oxide/Ag tunnel junctions with different thickness, *t*, of the NbN/oxide layer, in different magnetic fields applied parallel to the film plane. The $T_c$ of the NbN film, defined as the temperature where its resistance drops below 0.05% of the normal state value is shown for each device. For these tunnel junctions the oxide tunnelling barrier is formed by oxidising the NbN surface in air. Measurements are done at the base temperature of our cryostat, 0.29 K. Depending on the time of oxidation, we obtained tunnel junction resistance varying from 1 Ohm to few tens of Ohms. In zero field, all tunnel junction showed sharp coherence peaks and a well-formed gap characteristic of conventional Bardeen-Cooper-Schrieffer (BCS) superconductor. The film plane is precisely aligned parallel to the magnetic field within an accuracy of $\pm 1\%$ by recording the tunnel junction resistance for a fixed current at successive values of the sample rotator angle and choosing the angle for which the resistance maximum (For details see Appendix A). Under application of magnetic field, the tunnelling spectra broaden up to a maximum field of 105 kOe for *t* = 40 nm and 20 nm. However, signature of Zeeman splitting appears at the spectrum taken at 105 kOe for *t* = 10 nm. For *t* = 5 nm, Zeeman splitting become visible 60 kOe and is prominent at 105 kOe.

*Table 1*: Best fit parameters for NbN/oxide/Ag junctions with different thickness of NbN/oxide layer, *t*, measured at $H = 105$ kOe, $T = 0.29$ K ( shown in Figure 4(a) ).

| *t* (nm) | $T_c$ (K) | Δ (meV) | Γ (meV) | ζ | b | $G_0$ |
|---|---|---|---|---|---|---|
| 40 | 15.8 | 2.20 | 0.15 | 0.12 | 0.27 | 0.00 |
| 20 | 14.5 | 2.20 | 0.13 | 0.11 | 0.26 | 0.00 |
| 10 | 13.5 | 2.17 | 0.082 | 0.085 | 0.24 | 0.04 |
| 5 | 12.4 | 2.09 | 0.045 | 0.048 | 0.23 | 0.06 |

III.B. *Fitting the spin split density of states*

We start with the expression for tunnelling conductance for a spin split band,

$$G(V) = \frac{dI}{dV} \propto N_\uparrow |M_\uparrow|^2 \int_{-\infty}^{+\infty} \rho_\uparrow(E,H) \frac{\partial f(E+eV)}{\partial V} dE + N_\downarrow |M_\downarrow|^2 \int_{-\infty}^{+\infty} \rho_\downarrow(E,H) \frac{\partial f(E+eV)}{\partial V} dE \quad (1)$$

where $M_{\uparrow(\downarrow)}$ is the spin up(down) matrix element for transmission, $N_{\uparrow(\downarrow)}$ is the spin up(down) DOS of the normal electrode at Fermi level $E_F$, $\rho_{\uparrow(\downarrow)}$ is the spin up(down) superconducting density of states, and $f(E)$ is the Fermi function, where the energy is measured with respect to $E_F$. For a SIN junction, in zero magnetic field $N_\uparrow |M_\uparrow|^2 = N_\downarrow |M_\downarrow|^2$ and $\rho_\uparrow(E, 0) = \rho_\downarrow(E, 0)$, and we recover the usual expression for S/I/N tunnelling by substituting the BCS expression for the total density of states $\rho(E) = \rho_\uparrow(E, 0) + \rho_\downarrow(E, 0) = Re\left(\frac{|E + i\Gamma|}{\sqrt{(E + i\Gamma)^2 - \Delta^2}}\right)$. The imaginary component, $\Gamma$, is added to the quasiparticle free energy to account for the finite recombination lifetime[16]. For a ferromagnet the spin polarisation is given by,

$$P = \frac{N_\uparrow |M_\uparrow|^2 - N_\downarrow |M_\downarrow|^2}{N_\uparrow |M_\uparrow|^2 + N_\downarrow |M_\downarrow|^2}. \qquad (2)$$

*Table 2*: Best fit parameters for 5nm NbN/ Oxide/ Ag tunnel junction at 0.29K shown in Fig. 3(a).

| H (kOe) | Δ (meV) | Γ (meV) | ζ | b | $G_0$ |
|---|---|---|---|---|---|
| 0 | 2.18 | 0.105 | 0.011 | 0.23 | 0.00 |
| 30 | 2.16 | 0.095 | 0.017 | 0.23 | 0.00 |
| 60 | 2.13 | 0.085 | 0.024 | 0.23 | 0.05 |
| 90 | 2.11 | 0.065 | 0.038 | 0.23 | 0.05 |
| 105 | 2.09 | 0.045 | 0.048 | 0.23 | 0.06 |

While exact expressions for $M_{\uparrow\downarrow}$ are not available, they are essentially constant[17] as a function of energy over the few millivolts of interest here, as are $N_{\uparrow\downarrow}$. Rewriting equation (2) in terms of P we get,

$$G(V) = \frac{dI}{dv} \propto \frac{1 + P}{2} \int_{-\infty}^{+\infty} \rho_\uparrow(E, H) \frac{\partial f(E + eV)}{\partial V} dE$$

$$+ \frac{1 - P}{2} \int_{-\infty}^{+\infty} \rho_\downarrow(E, H) \frac{\partial f(E + eV)}{\partial V} dE \quad (3)$$

To calculate $\rho_\uparrow(E, H)$ and $\rho_\downarrow(E, H)$ we use the expressions from Maki's theory[18] which includes the effects of orbital de-pairing, the Zeeman splitting of the spin states, and spin-orbit scattering,

$$\rho_{\uparrow\downarrow}(E) = \frac{\rho(0)}{2} sgn(E) Re\left(\frac{u_\pm}{(u_\pm^2 - 1)^{\frac{1}{2}}}\right), \quad (4a)$$

where $u_\pm$ are defined by the self-consistent set of two coupled equations,

$$u_\pm = \frac{E + i\Gamma \mp \frac{\mu_B H}{1+G_0}}{\Delta} + \frac{\zeta u_\pm}{\left(1 - u_\pm^2\right)^{1/2}} + b\left(\frac{u_\mp - u_\pm}{\left(1 - u_\mp^2\right)^{1/2}}\right); \quad (4b)$$

$\mu_B$ stands for Bohr magneton and $\zeta$ and $b$ are two parameters that account for orbital depairing and spin-orbit scattering respectively. In addition, we include two addition parameters; the broadening parameter $\Gamma$ that we have described before and a Fermi liquid parameter[19] $G_0$ which becomes significant only close to $H_{c2}$ (or $T_c$). Eq.(4b) is not easy to solve. Numerical methods to try to directly solve the two coupled equations often fail to converge. The most powerful method is the one outlined developed by Alexander et al[19], where Eq.(4b) is recast into 4 equations with complex unknown. The benefit of using these four equations instead of Eq. (4b) is that for all choices of parameters, the Newton-Raphson method converges for wide range of parameters and starting values.

In Fig. 3(a) we show the fit of the tunnelling spectra taken at various magnetic field for $t = 5\ nm$ with equations 3 and 4 (with P=0). The fit parameters are given in Table 1. $\zeta$ increases with field following the relation[20], $\zeta(H) = \zeta(0) + AH^2$ (Fig. 3(b)), reflecting the increase in orbital depairing with increasing magnetic field. The quadratic variation of $\zeta$ with magnetic field is theoretically expected when the electronic mean free path[21], $l \ll \xi$, which is the case here. The residual value $\zeta(0)$ on the other hand, suggests that some mechanism of pair-breaking is present even in zero magnetic field. Even though the origin of this is unknown, this has also been observed before in superconducting Al based tunnel junction.

In Fig. 4(a), we fit the tunnelling spectra at 105 kOe, 0.29 K for devices with different $t$ (fit parameters in Table 2). The spin-orbit coupling parameter, $b$, displays a weak increasing trend with increasing $t$. On the other hand, $\zeta$ increases rapidly at low thickness and tends to saturate asymptotically towards the bulk value at higher thickness.

III.C. *Measurement of spin polarisation in SIF junctions*

*Table 3*: Best fit parameters for various *H* for 5nm NbN/oxide/Co junction shown in Figure 5 at T = 0.78K. The tunnel barrier is formed by oxidising NbN surface in air.

| H (kOe) | Δ (meV) | Γ (meV) | ζ | b | $G_0$ | P |
|---|---|---|---|---|---|---|
| 0 | 2.055 | 0.175 | 0.007 | 0.25 | 0 | 0 |
| 60 | 1.995 | 0.162 | 0.014 | 0.28 | 0.02 | 0 |
| 90 | 1.98 | 0.145 | 0.029 | 0.28 | 0.04 | 0.21 |
| 105 | 1.97 | 0.035 | 0.035 | 0.28 | 0.06 | 0.21 |

We now investigate the suitability of NbN/oxide based (NbN oxidised in Air) SIF junctions to extract the spin polarisation of Co. For this we use the NbN/oxide layer thickness *t* = 5 nm, and deposit an 18 nm thick layer of Co as the counter-electrode, subsequently capped by depositing a 15 nm Au overlayer. Fig. 5 shows the tunnelling spectra of NbN/Oxide/Co junction measured at 0.78 K in different magnetic field. In this case we had to use larger current due to low resistance of the tunnel junction resulting in a larger contact heating that prevented us from acquiring the tunnelling spectra at the base temperature of the cryostat. Nevertheless, at 90 and 105 kOe we observe clear splitting and a pronounced asymmetry between positive and negative bias voltage arising from the spin polarisation of Co. We fit these two spectra using equations (3) and (4) and obtain a spin polarisation, $P \approx 0.21 \pm 0.005$. We can also fit the spectrum taken at 60 kOe using the same value of spin polarisation even though the error on *P* is much larger for lower fields. Our obtained value of spin polarisation for Co is consistent with spin polarised photoemission measurement[6] and somewhat smaller than the value $P \sim 0.35$ reported in early measurements by Meservey and Tedrow[4]. However, in that paper a simplified scheme that ignored spin-orbit coupling was used to analyse the data, and the spin polarisation was obtained only from the conductance values at the spin up and spin down coherence peaks. If we use the same scheme, we obtain a $P \approx 0.36$, consistent with that report. However, it has been shown that any analysis that ignores spin-orbit coupling can introduce very large errors in the spin polarisation values and is not reliable.

So far, all tunnel junctions that we discussed were made by oxidising the NbN in air. One problem with this scheme, while we obtained good tunnelling characteristics the resistance of most tunnel junctions were very small ( < 0.5 Ohm) , thus requiring higher maximum currents in *I-V* measurements. This is a problem in sample in vacuum cryostats where large currents result in significant contact heating. While the resistance of the junction could in-principle be increased by increasing the thickness of the barrier by oxidising for a longer time, this severely deteriorated the tunnelling characteristics for NbN/oxide/Co junctions. This degradation is normally due to the formation of defects in the tunnel barrier which acts as traps for the electron, where it loses its energy before tunnelling to the other electrode. To circumvent this problem, we attempted to oxidise in 99.99% pure oxygen instead of air. In this case, we could routinely get NbN/Oxide/Co tunnel junctions with normal state resistance in the range 20-30 ohms. The tunnelling spectra recorded at 0.29 K for one such junction is shown at in Fig. 6(a) for different magnetic field. In Fig. 6(b) we show the theoretical fit of the 105 kOe spectrum. While we need to use significantly larger value of Γ to fit the data (see Table 3), the spin polarization value remains unaltered at $P \approx 0.21$. In Figure 6(c) we show the normalised tunnelling spectrum at 105 kOe for a higher temperature 1.57 K. The asymmetry in the spectrum is still visible at this temperature and the fit converges at the same value of *P*. We also found these tunnel junctions to have much higher reproducibility than junctions oxidised in air.

*Table 4*: Best fit parameters of 5 nm NbN/oxide/Co junction measured at 105 kOe, shown in Figure 6(b).

| T (K) | Δ (meV) | Γ (meV) | ζ | b | $G_0$ | P |
|---|---|---|---|---|---|---|
| 0.29 | 1.76 | 0.24 | 0.05 | 0.23 | 0.03 | 0.21 |
| 1.57 | 1.76 | 0.26 | 0.035 | 0.23 | 0 | 0.21 |

**IV. Conclusion**

In conclusion, we have shown that superconducting NbN with oxidised surface layer is a viable and simple alternative to Al for measuring spin polarisation of ferromagnets using Meservey-Tedrow method by fabricating SIF junctions. In contrast to Al, for which measurements need to be performed at temperatures well below 1 K, NbN offer the possibility of performing these measurements in conventional $^4$He cryostats. Combined with the ease of fabricating these junctions without any sophisticated lithography, we believe this will open the possibility of wider applicability of this technique to measure spin-polarisation in new materials.

**Appendix A: Aligning the film plane parallel to magnetic field**

To obtain spin-splitting of the superconducting density of states, one of the crucial requirements is to make the film plane parallel to the magnetic field for minimizing the orbital currents. After applying a magnetic field between 20-30 kOe, we send a constant current ( 100-200 μA ) so that the voltage developed at the junction is close to the coherence peak voltage. We then we rotate the sample holder by the rotator and find the angle that maximizes the voltage across the junction. This corresponds to the aligned orientation. This is shown in Figure 7(a) where we follow the convention that $H\|$(film plane) corresponds to $\theta = 0$. It is easy to see from the *I-V* curves recorded at different $\theta$ (Fig. 7(b)) that the voltage across the junction at a constant current will decrease with increasing $\theta$. In Fig. 7(c), we plot the tunnelling spectra at 0.32 K, 105 kOe for different values of $\theta$. As expected, the most pronounced spin-splitting appears for the spectrum at $\theta = 0$.

*Acknowledgement:* We thank Rishabh Duhan, Soumyajit Mandal and Vivas Bagwe for help with experiments, and Gaurav Agarwal for his initial involvement in writing some parts of the codes used to analyse the data. This work was supported by the Department of Atomic Energy, Govt. of India (Grant No. 12-R&D-TFR-5.10-0100).

The authors declare no competing financial interest.

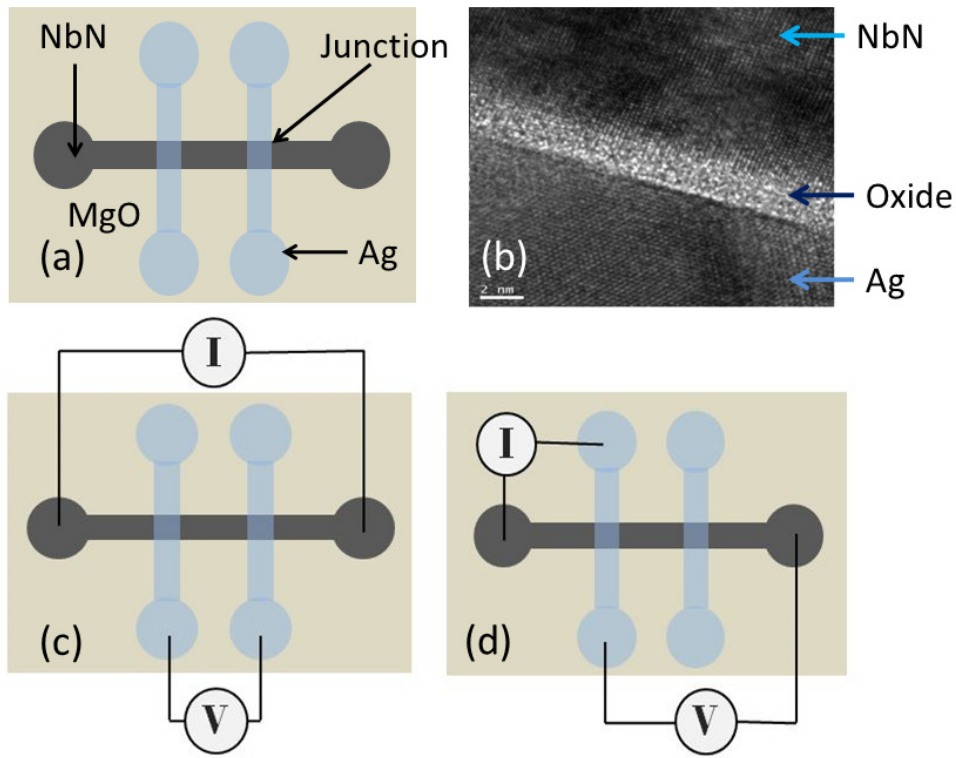

**Figure 1.**

(a) Schematic diagram of the device. 300 μm wide NbN strip at the bottom (dark grey) and two 300 μm wide cross-strips at the top (light blue), making for two junctions of cross-sectional area 300×300 μm².
(b) High resolution transmission electron image of NbN/Oxide/Ag junction showing 2 nm thick oxide barrier separating NbN and Ag.
(c) Configuration of 4 probe transport measurement to measure the resistance of NbN stripe.
(d) Configuration of 4 probe transport measurement to measure the *I-V* characteristics of the Junction.

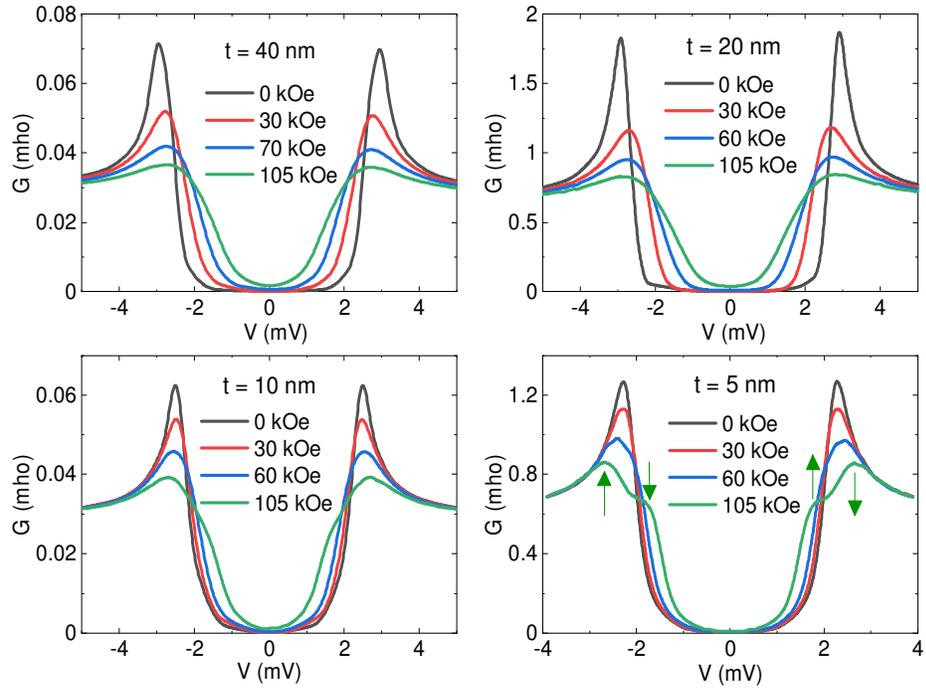

**Figure 2.**

*G(V) vs V* tunnelling conductance spectra for NbN/Oxide/Ag junctions at $T = 0.29$K in different magnetic fields. The four panels correspond to different thickness of NbN/oxide layer, *t*, used while making the tunnel junction. Splitting for up(down) spin sub-bands is indicated in the figure by up(down) arrows for *t* = 5nm tunnel junction.

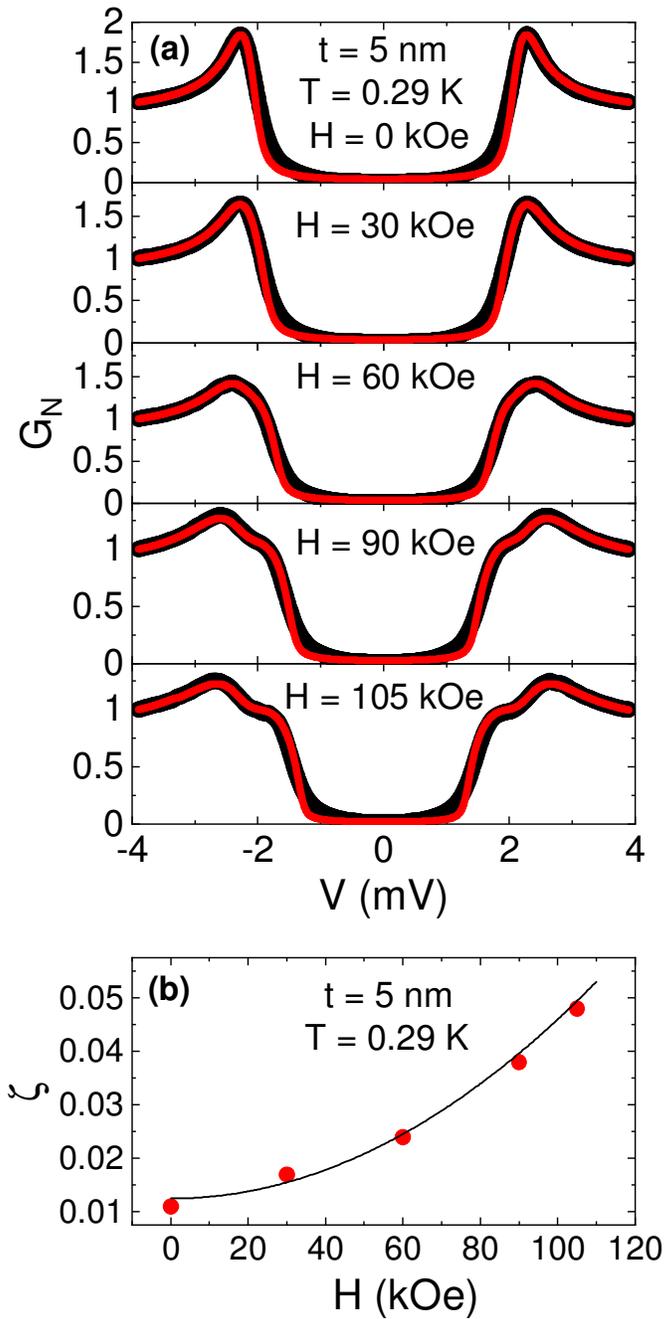

**Figure 3.**

(a) Normalised tunnelling conductance spectra ($G_N(V)$ vs $V$) (black points) along with theoretical fits (red line) for $t = 5$ nm NbN/Oxide/Ag junction measured at T = 0.29K in different magnetic fields. The spectra are normalised at 4 mV.

(b) Depairing parameter $\zeta$ extracted from the fits plotted with magnetic field H and fitted with $\zeta(H) = \zeta(0) + AH^2$ to show the $H^2$ dependence of $\zeta$; The best fit parameters are $\zeta(0) = 0.0125$ and $A = 1.522 \times 10^{-4}$ /kOe$^2$.

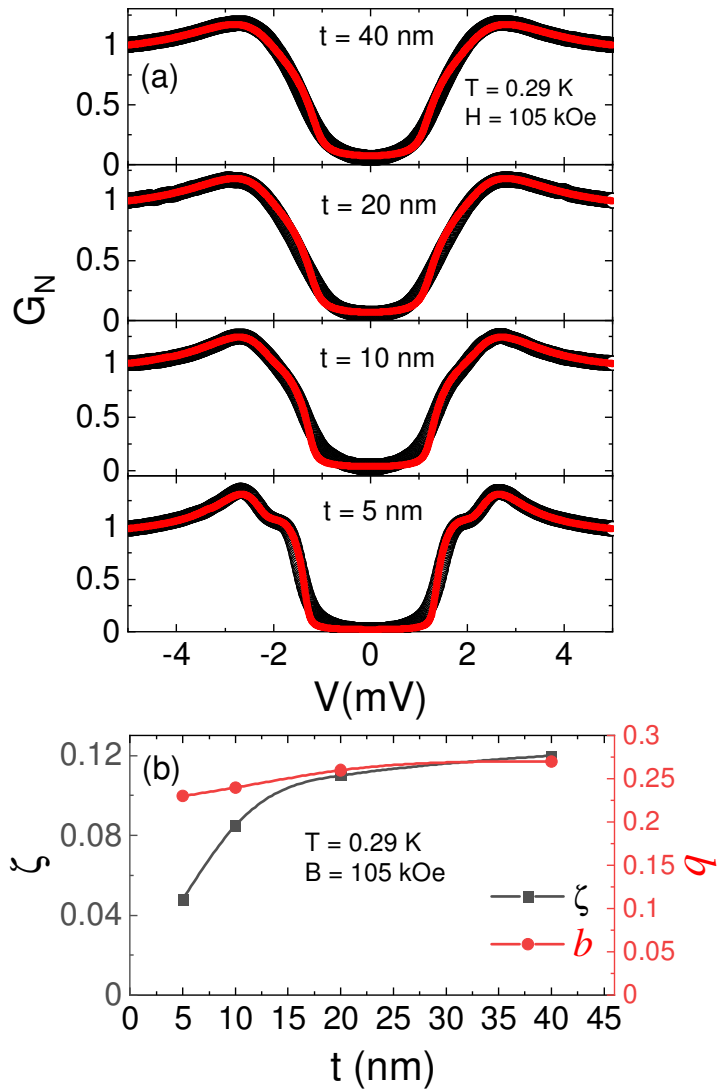

**Figure 4.**

- (a) Normalised tunnelling conductance spectra ($G_N(V)$ vs $V$) (black points) at H = 105 kOe, T = 0.29K along with theoretical fits (red line) for NbN/Oxide/Ag tunnel junctions with $t$ = 40 nm, 20 nm, 10 nm, 5nm junction. The spectra are normalised at 5 mV.
- (b) Variation of ζ and b with the thickness of NbN/oxide layer, $t$, at H= 105 kOe and T = 0.29 K.

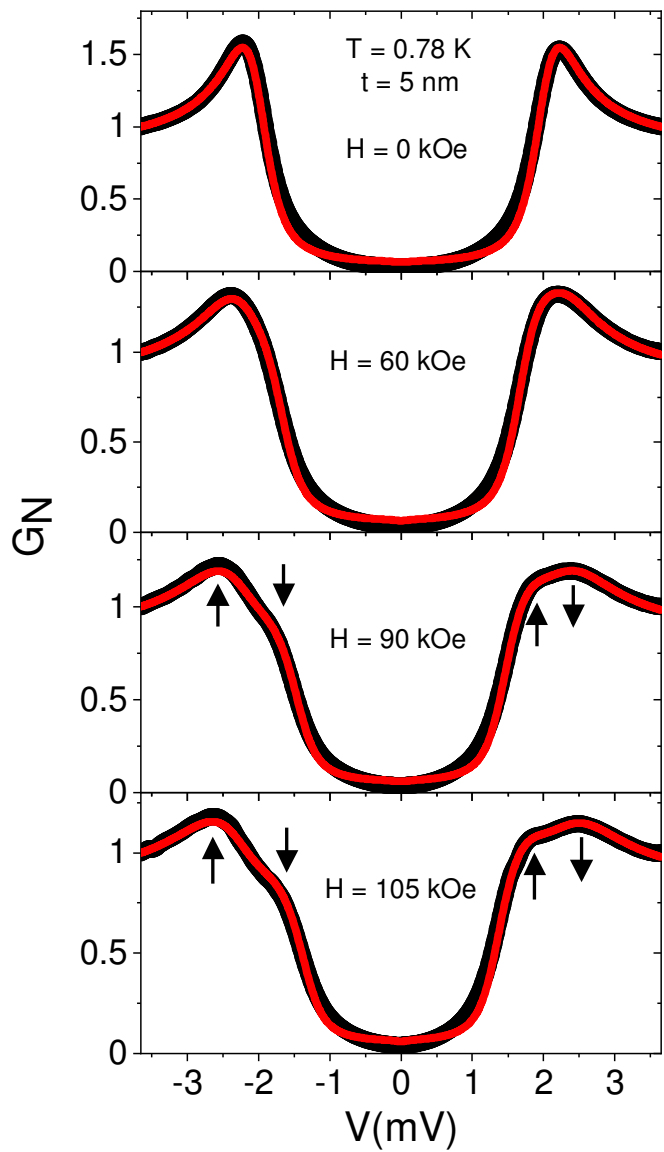

**Figure 5.**

Normalised tunnelling conductance spectra ($G_N(V)$ vs $V$) (black points) along with theoretical fits (red line) for $t = 5$ nm NbN/Oxide/Co junction at $T = 0.78$ K and H = 0 kOe, 60 kOe, 90 kOe, 105 kOe. Splitting for up(down) spin sub-bands is indicated in the figure by up(down) arrows.

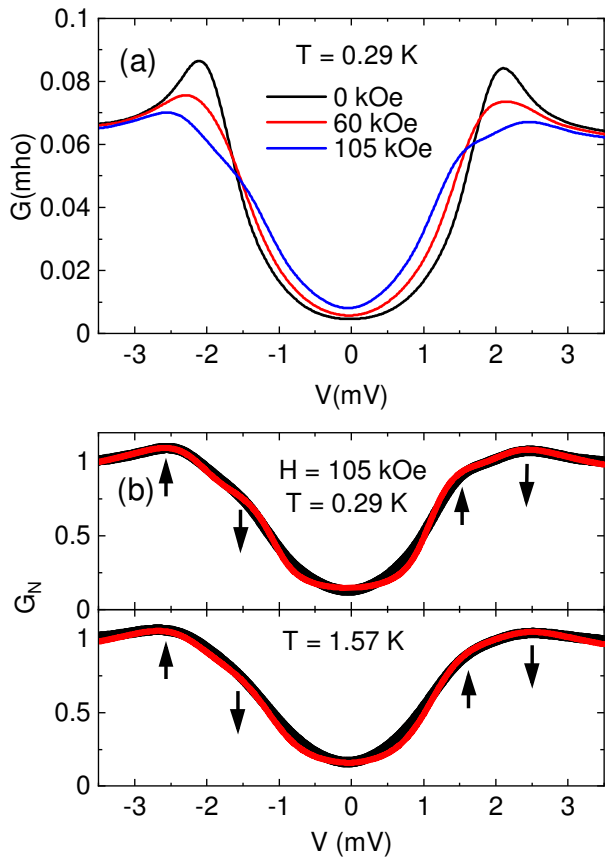

**Figure 6.**

(a) *G(V) vs V* tunnelling conductance spectra at *T* = 0.29 K for *t* = 5 nm NbN/Oxide/Co junction for which the oxide barrier is created in high purity oxygen.

(b) Normalised tunnelling conductance spectra (black point) along with the fits (red line) for the spectra recorded in 105 kOe at temperatures *T* = 0.29 K (*upper panel)* and *T* = 1.57 K (*lower panel*). Splitting for up(down) spin sub-bands is indicated in the figure by up(down) arrows.

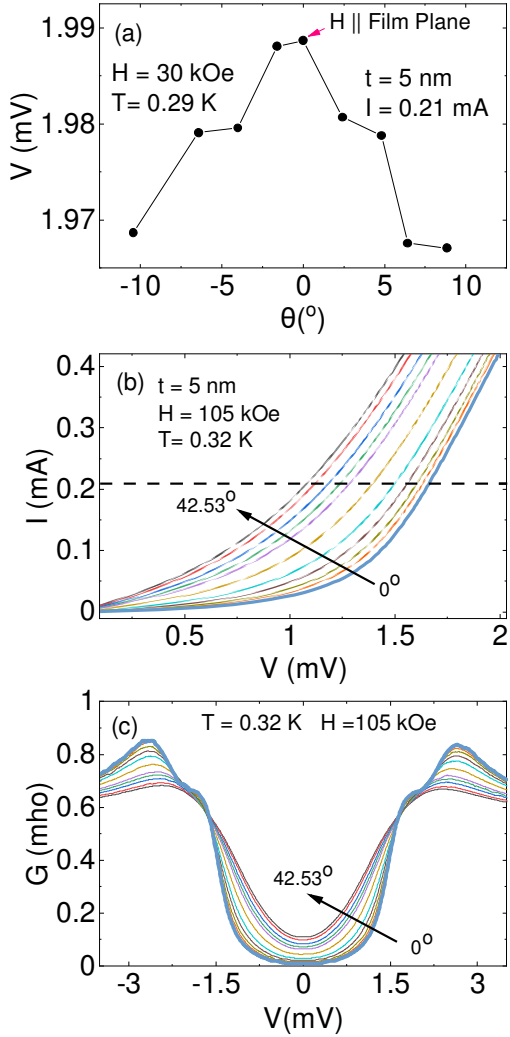

**Figure 7.**

(a) Junction voltage (V) for a fixed current, $I = 0.21$ mA, as a function of the angle, $\theta$, between the film plane and the magnetic field for the NbN/Oxide/Ag junction with $t = 5$ nm. The applied magnetic field, $H = 30$ kOe and the measurement is done at 0.29 K. $\theta = 0$ corresponds to the orientation where $H$ is parallel to the film plane.

(b) Expanded view of the *I-V* characteristics of the same junction at for $\theta$ for $H = 105$ kOe and $T = 0.32$ K. The dashed line corresponds to the voltage cut at constant current of 0.21 mA. The successive values of $\theta$ going along the arrow are, $42.53^0$, $36.11^0$, $29.70^0$, $25.68^0$, $23.27^0$, $16.85^0$, $10.43^0$, $8.83^0$, $4.01^0$, $2.41^0$ and $0^0$. The $\theta = 0$ curve is highlighted with thick light blue line.

(c) *G(V)* vs *V* tunnelling conductance spectra for variation of the angle of rotation ($\theta$) with respect to the applied magnetic field direction at $T = 0.32$ K for $H = 105$ kOe. The successive values of $\theta$ going along the arrow are, $42.53^0$, $36.11^0$, $29.70^0$, $25.68^0$, $23.27^0$, $16.85^0$, $10.43^0$, $8.83^0$, $4.01^0$, $2.41^0$ and $0^0$. The $\theta = 0$ curve is highlighted with thick light blue line.